\documentclass[12pt]{article}
\usepackage{amsfonts,amsmath,amscd,amssymb,amsthm}

\newfont{\eufm}{eufm10 scaled\magstep1}

\begin{document}

\title{Waveless Approximation Theories of Gravity}

\author{James A. Isenberg\\
 Department of Physics and Astronomy\\
 University of Maryland\\
 College Park, Maryland 20742}
 
 \date{May, 1978}
 
 \maketitle

\textbf{Note from the Author}\\

During the 1970's, there was a lot of interest in developing good numerical simulations of the production of gravitational radiation by strongly interacting astrophysical systems (the archetype being a binary pair of black holes). As a graduate student at the University of Maryland during  that time, with strong encouragement from my advisor, Charles Misner, I developed a collection of approximation schemes which were designed to try to avoid the fatal numerical instabilities which arose during the course of all of the contemporary attempts to evolve the full system of Einstein's equations. 
The idea of these schemes was to mathematically decouple the ``gravitational wave" production from the evolution of the matter and the slowly changing ``induced gravitational fields" of the astrophysical systems. The equations governing the evolution of the matter and induced gravitational field system were designed to be a form of enhanced Newton-Euler system: The matter fields would evolve via Euler-type equations, while the induced gravitational fields would be determined by a system of elliptic equations to be solved for certain pieces of the metric. These equations were set up to ignore gravitational radiation, except for the possible inclusion of terms accounting for the loss of energy due to wave propagation. To determine the gravitational radiation production, the idea was to incorporate the motion of the matter and induced gravitational fields into source terms for a system of linear wave equations.

In 1978, I wrote a paper which describes some of the equation systems I developed for the matter and induced gravitational field evolution. I called these systems "Waveless Approximation Theories of Gravity", and I submitted a paper with this title to Physical Review D. It was rejected by the referees, based on the reasonable contention that I hadn't tried to numerically implement the ideas. Not being any good at numerical work, I lay the paper and the ideas aside, and I went on to  other things. The only record of it which I left was a brief mention towards the end of an article I wrote with Jim Nester for the Einstein Centenary volume "General Relativity and Gravitation" [edited by A. Held, Plenum, 1980].

A number of years later, John Friedman noticed that one of the popular methods developed by J. R. Matthews and James Wilson to study gravitational radiation production (It is sometimes called the CFC or ``Conformal Flatness Condition"  approach) is very similar to one of my Waveless Approximation Theories. (John happened to have read my article with Nester.) He has been kind enough to suggest that people call this the Isenberg-Matthews-Wilson method, but has noted that there was no real reference to my work available. To remedy this, he suggested that I post the 1978 article. I am doing this here. 

The version appearing here is unchanged from what I wrote in 1978. In reading it over, I have found no serious errors, except for the claim that the equations constituting ``WAT-II" are elliptic, and consequently always admit a unique solution. For general fields, they are not. I hope people find this article useful.

\clearpage

\begin{abstract}
The analysis of a general multibody physical system governed by
Einstein's equations in quite difficult, even if numerical methods
(on a computer) are used. Some of the difficulties -- many coupled
degrees of freedom, dynamic instability -- are associated with the
presence of gravitational waves.  We have developed a number of
``waveless approximation theories'' (WAT) which repress the
gravitational radiation and thereby simplify the analysis. The
matter, according to these theories, evolves dynamically.  The
gravitational field, however, is determined at each time step by a
set of elliptic equations with matter sources. There is reason to
believe that for many physical systems, the WAT-generated system
evolution is a very accurate approximation to that generated by
the full Einstein theory.
\end{abstract}

\pagebreak

  \setcounter{equation}{0}

 \indent According to Einstein's theory, most
nonspherical, nonstationary multibody systems must radiate
gravitationally [1]. Whether or not one is interested in the
gravitational waves themselves, they make the analysis of the
evolution of such a system quite difficult to carry out. Numerical
computers can overcome some of these difficulties, such as those
associated with many coupled degrees of freedom, nonlinearity, and
gauge coordinate freedom; but the hyperbolic nature of Einstein's
equations (including gravity waves) leads to problems of numerical
stability.

\indent While the gravitational waves make it hard to analyze the
motion of a multibody system, the studies of Thorne [2], Smarr [3]
and others indicate that in some cases, the real physical effects
of the waves upon the motion of the bodies (``radiation back
reaction'') may be quite small.  We are thus motivated to develop
an approximation to Einstein's theory (coupled to matter) which
neglects radiation in calculating the motion of a gravitating
system of bodies.  Such a computational tool could be useful not
only in studies which ignore gravity waves, but also in studies
which seek to find the radiation emitted by a system:  A waveless
approximation theory (``WAT'') could be used first, to determine
the motion of the bodies; then one could compute what waves are
emitted by bodies so moving.  A WAT would effectively decouple and
thereby simplify the analysis of gravity wave production.

\indent No one knows exactly how to identify the ``radiation
terms'' in Einstein's theory.  Thus there is no clearcut, unique
way to neglect radiation in Einstein's theory.  We have therefore
developed a few waveless approximation theories.  Some of them are
formally superior, some are easier to use in doing calculations,
and some appear to give good approximations for a wider class of
systems.  Much testing of the various versions of WAT remains to
be done before we can say which of the WAT's holds the most
practical promise.  It may happen that more than one version can
be used profitably.

\indent The original (and in some ways the best) ``waveless
approximation to Einstein's theory'' is the Newtonian theory of
gravity.  When supplemented by the Euler equations for fluid
motion [4], Newton's theory determines the evolution of a
multibody (continuous fluid) self gravitating system as follows:
The matter is described by the mass-density $\rho(x,t)$
the 3-velocity $v^q(x,t)$, and the pressure $p(x,t)$.  [The
pressure is assumed to be some function of $\rho$, according to a
prescribed equation of state.]  The gravity is described by a
single potential $\Phi(x,t)$.  Only the matter variables $\rho$
and $v^\alpha$ are dynamic.  They change in time according to the
equations [5].
 \begin{equation}
\dot{\rho} = -\partial_a (p v^a)
\end{equation}
and
\begin{equation}
\dot{v}^a = - v^m \partial_m v^a - \frac{1}{p} \; \partial^a p +
\partial^a \Phi .
\end{equation}
As the matter moves about, the gravitational potential $\Phi$
slavishly follows, as determined by the elliptic equation
\begin{equation}
\partial^2 \Phi = - 4 \pi \rho.
\end{equation}
Gravity has no dynamics of its own, in the Newtonian theory.

\indent For many systems, (most of those we may encounter in the
solar system) the Newtonian theory provides a very accurate
description of the motion.  And it is very easy to work with.
[Indeed, the gravitational potential equation (3) can always be
solved, in integral form, thereby eliminating $\Phi$ explicitly
from the matter evolution equations (1) and (2).]  However at
relativistic speeds $(v^2 \rightarrow 1)$ and relativistic
densities $(p \rightarrow 10^{15} gm/cc)$ the Newtonian
description becomes progressively inaccurate.

\indent Einstein's theory, which agrees with present-day
experiments to a high degree of precision, replaces the single
gravitational potential by the ten components of the metric.  And
it replaces the single linear elliptic equation (3) by the
nonlinear, mixed elliptic-hyperbolic set of Einstein's equations.
Broken down into 3 + 1 dynamic form (convenient for later
discussion), the Einstein system with fluid source takes the
following form:  The field variables are the metric
\begin{equation}
g = -N^2 dt^2 + \gamma_{ab}(dx^a + M^a dt)(dx^b + M^b dt),
\end{equation}
the extrinsic curvature
\begin{equation}
K_{ab} = - 2N \mathcal{L}_{+}(\gamma_{ab})
\end{equation}
and the fluid stress energy tensor
\begin{equation}
T^{\alpha,\beta} = \rho U^\alpha U^\beta + pg^{\alpha\beta}.
\end{equation}
The lapse $N$ and shift $M^\alpha$ are completely arbitrary
throughout time.  Four of Einstein's equations,
\begin{equation}
0 = R + (tr K)^2 - K^m_n K^n_m - 16 \pi[(U^\bot)^2 \rho - p]
\end{equation}
and
\begin{equation}
0 = \nabla_c K^c_b - \nabla_b(tr K) - 8 \pi \rho U_a U^\bot
\end{equation}
are constraint equations for the initial data.  These constraints
can be made explicitly elliptic, using the
conformal-transverse-trace decomposition scheme of York [6].  The
rest of the system consists of time evolution equations:
\begin{equation}
\dot{\gamma}_{ab} = - 2 NK_{ab} + \mathcal{L}_M \gamma_{ab},
\end{equation}
\begin{eqnarray}
\dot{K}^a_n &=& N\{ R^a_n + tr K K^a_n - [p \delta^a_n + \rho U^a
U_n] \nonumber \\
 & & - \frac{1}{2} \delta^a_n [\rho(U^\bot)^2 - 3p]\} - \nabla^a
 \nabla_n N + \frac{1}{N} \mathcal{L}_M K^a,
\end{eqnarray}
\begin{equation}
\dot{\rho} = \mathcal{F}[N, M, \gamma, K; \rho, U, p],
\end{equation}
and
\begin{equation}
\dot{U}^m = \mathcal{F}^m[N,M,\gamma,K; \rho, U, p].
\end{equation}
Here $\mathcal{F}$ and $\mathcal{F}^m$ are easily calculated (from
the action) functionals of the indicated variables.  Despite the
elliptic initial value constraints (7) and (8), this set of
equations is predominantly a dynamic (hyperbolic) set.  The
gravitational field does not slavishly follow the matter; it
evolves on its own.  This contributes to computational
instability.  This instability, together with strong coupling and
nonlinearities evident in (7)- (12), make the full Einstein
theory very difficult to use for computation, even on a large
computer.

\indent The waveless approximation theories are designed to
compromise between the Newtonian and Einsteinian theories.  We
would like as much of the gravitational fields as possible to be
nondynamic - to be determined elliptically at each time by the
evolving matter.  We would like the nonlinearities and coupling in
the equations to be cut down.  At the same time, we want the
WAT-generated motion of a given system to be as close as possible
to that computed using Einstein's theory.

\indent  WAT-I, which leans slightly towards the Newtonian side of
the compromise, is the best motivated among all the WAT's.  It is
founded on the notion that the conformal 3-geometry on a maximal
$(tr K = 0)$ spacelike slice is closely associated to
gravitational radiation [7].  Thus WAT-I produces maximal
spacelike slices only, and it represses the intrinsic conformal
curvature on these slices.

\indent We derive the WAT-I field equations by building intrinsic
conformal flatness and maximal slicing directly into the
Lagrangian action of Einstein:
\begin{equation}
 S = \int [^4 R + L_n]_\eta
 \end{equation}
 [Here $^4R$ is the spacetime scalar curvature, $L_M$ is the
 matter Lagrangian, and $\eta$ is the volume element].  Thus we
 compute $^4R$ out of the metric
\begin{equation}
g = -N^2 dt^2 + e^{2 \psi} \delta_{ab}(dx^a + M^a dt)(dx^b + M^b
dt)
\end{equation}
[where $\delta_{ab}$ is the flat metric] and we require that the
momentum conjugate to $\psi$ must vanish.  The matter Lagrangian
is chosen as in Einstein's theory, but with the metric taking the
specialized form (14).  Varying the action (13), with perfect
fluid source fields [8], we obtain the following set of field
equations for the five gravitational potentials $(N, M^0, \psi)$
and for the matter fields $\rho, U^\alpha, p)$:
\begin{equation}
\partial^2 \psi + (\partial_m \psi)(\partial^m \psi) = \frac{e^{2\psi}}{8N^a}
 [(\ell M)^a_b (\ell M)^b_a] + 2 \pi[\rho (U^\alpha)^2 - p]
\end{equation}
\begin{equation}
\partial^2 N + (\partial_m N) \frac{(\partial_m \psi)}{2} =
\frac{1}{2N} e^{2 \psi}[(\ell M)^a_b (\ell M)^b_a] +
N\left(\frac{3}{2}p - \rho\right),
\end{equation}
\begin{equation}
\partial_b\left[\frac{e^{2\psi}}{N} (\ell M)^b_a \right] = 8 \pi U_f U^\bot
\rho,
\end{equation}
\begin{equation}
\dot{\rho} = \mathcal{F}[N, M, \psi; \rho, U, p]
\end{equation}
\begin{equation}
\dot{u}^m = \mathcal{F}^m[N,M, \psi; p, U, p]
\end{equation}
Here $(\ell M)^a_b = \partial^a M_b + \partial_b M^0 - \frac{2}{3}
\delta^a_b \partial_m M^m$ (conformal killing operator in flat
space).
\\
\indent Clearly the gravitational fields of WAT-I are of the desired
``slave'' type.  That is, while the matter evolves according to
(18) and (19), the potentials $N, M^a, \psi$ follow it around, as
determined by the five elliptic equations (15)-(17) and the
accompanying boundary value problem.  The equations (17)-(19) are
coupled and non-linear.  However, the differential operators
appearing in these equations are all strongly elliptic and self
adjoint [9].  So while the WAT-I boundary value problem is nowhere
as simple as that of Newton's theory, it seems to admit unique
solutions for any chosen matter fields $\rho, U^\alpha, p(p)$
[10].  Moreover, the nonlinearities and coupling in these
equations are relatively mild.  We see this especially if we
consider matter fields and boundary conditions with planar
symmetry, in which case we can always obtain explicit analytic
solutions for $(N, M^a, \psi)$ [11].  More generally, while we
can't always find analytic solutions, the boundary value problem
for WAT-I is amenable to numerical solution, without any
particular inherent danger of numerical instability.

\indent How accurately does the WAT-I treatment of a given
multibody system approximate the Einstein treatment of the same
system (as specified by free initial data)?  We first note that,
unlike Newtonian solutions, WAT-I solutions can satisfy the full
set of Einstein's equations.  A prime example is the Schwarzschild
solution.  This can happen because the five WAT-I gravitational
equations are a subset of the Einstein equations (7)-(10) for a
maximal, intrinsically conformally flat, slicing.  [Such is not
the case for the Newtonian equation (3).]  A means of gauging the
inaccuracy of the WAT-I prescribed motion for a given system now
suggests itself:  Consider the collection of terms in the Einstein's equations
which are not included in the WAT system.  They are primarily the
matter and curvature terms appearing in the tracefree right hand
side of eq. (10) which prevent a set of intrinsically conformally flat,
maximal initial data from retaining these properties as it
evolves.  So it is these terms which cause the WAT-I evolution and
the Einstein evolution to diverge.  Thus if we monitor these terms
as we carry out a WAT-I analysis of a given multibody system, we
get some idea of how different an Einstein analysis of the same
system would be.

\indent Another means for comparing WAT-I and Einstein treatments
of multibody systems, particularly those on an astrophysical
scale, is provided by the PPN scheme [12], along with its higher
order extensions [13].  Since the standard PPN spacetime metric
takes exactly the form (14) [i.e. it admits an intrinsically
conformally flat slicing], we see that to PPN order, WAT-I and the
Einstein theory are identical.  The two theories diverge at higher
(``PPPN'') orders, but this is not surprising, since gravitational
radiation is important in these higher order approximations.  We
can use the PPPN analyses to parametrize the inaccuracy of WAT-I,
as applied to systems of interest.

\indent It is not only gravitational radiation that is ruled out
when we require (as in WAT-I) that a spacetime admit a maximal,
intrinsically conformally flat metric.  The Kerr ``rotating black
hole'' spacetime is stationary [and therefore contains no
radiation], and yet it admits no such metric foliation [14].  We
would like a waveless approximation theory which could handle
Kerr, and other systems of rotating bodies.  Thus we look for a
WAT which permits intrinsic conformal curvature, and in fact
determines the intrinsic conformal geometry on every (maximal?)
slice elliptically.

\indent No way has been found to directly build into the action
principle an elliptic determination of the intrinsic conformal
geometry on each (maximal) spacelike slice.  If, however, we are
willing to sacrifice the advantages of an action [e.g., the
canonical formalism, and the Noether conservation laws], then we
can set up new waveless approximation theories which do include
non-dynamic conformal curvature.  We describe a few of them here.

\indent In some sense, the dynamics of the intrinsic conformal
geometry is contained in the transverse traceless part of the
extrinsic curvature tensor $K_{(TT)}$.  We are thus motivated to
kill both $K_{(TT)}$ and its time evolution.  The vanishing of
$K_{(TT)}$, together with the maximal slide condition, allows us
to replace $K^a_b$ by the conformal Killing form:
\begin{equation}
K^a_b = (LW)^a_b = \nabla^a W_b + \nabla_b W^a - \frac{2}{3}\;
\delta^a_b (\nabla \cdot W)
\end{equation}
The requirement that $K_{TT}$ vanish in the Einstein equations is
a complicated condition, involving inverse operators.  But if we
instead ask that the time derivative of the full traceless part of
$K^a_b$ vanish, then we get a manageable equation:
\begin{eqnarray}
0 = R^c_b - \frac{1}{3} \; \delta^c_b R &-&
\frac{1}{N^2}\;[\nabla^c N \nabla_b N - \frac{1}{3}\;
\delta^c_b \nabla^c N] \nonumber \\
& - & \rho[U^c U_b - \frac{1}{3}\; \delta^c_b U^m U_m] +
\frac{1}{N}\; \mathcal{L}_M(LW)^c_b
\end{eqnarray}
Examination of $R_{ab}$ as a differential operator on the
conformal metric $\widetilde{\gamma}_{ab}$ indicates that (21)
determines $\widetilde{\gamma}_{ab}$ elliptically.  We build
WAT-II by combining (21) with some means of determining the
conformal scale factor $\psi$, the lapse $N$, and the shift $M^a$.
The method for doing this which is most faithful to Einstein's
equations (and therefore most accurate) is to combine (20) and
(21) with the Einstein constraint equations, the matter evolution
equations (11)-(12), and finally the maximal slice and minimal
distortion conditions [15]:
\begin{equation}
0 = (tr \dot{K}) \Rightarrow 0 = R - \frac{\nabla^2 N}{N} -
\frac{3}{2}\; \rho(U_2)^2 - \frac{3}{2} \; p - \rho U^m U_m
\end{equation}
\begin{equation}
0 = (\widetilde{\gamma})^m_n \Rightarrow 0 = \nabla_m[LM]^m_n -
\nabla_m[N(LW)^m_n].
\end{equation}
Then WAT-II proceeds by (a) evolving the matter fields according
to (11) and (12), and (b) solving for the thirteen gravity
potentials $\widetilde{\gamma}_{ab}$, $\psi, N,M^a, W^a$ using (9),
(10), (21), (22), and (23).  The determination of the gravitational field is
apparently quite complicated, but since it is elliptic in nature,
WAT-II is still a simpler system than Einstein's equations for
numerical computation.

\indent A simpler version of WAT-II is obtained if we decouple the
determination of $\widetilde{\gamma}_{ab}$ from that of the other
gravitational potentials.  We can do this by using the equations
of WAT-I to find $N, M^a$ and $\psi$, setting $W^a = M^a$, and
then solving equation (21) for $\widetilde{\gamma}_{ab}$.  This
decoupled version of WAT-II involves much simpler equations, but
is likely to be much less accurate than the coupled version.

\indent WAT-III is obtained by forgetting all about WAT-I, [i.e.
forgetting conformal decomposition and maximal slicing] and
carrying the idea of WAT-II [requirement that the traceless part
of (10) vanish] one step further.  For WAT-III, we replace
$\dot{K}^a_b$ and $\dot{\gamma}_{ab}$ by zero everywhere in the
Einstein equations (7)-(10) and then attempt to solve what
remains.  Setting $\dot{\gamma}_{ab} = 0$ in (9) allows us to
replace $K_{ab}$ by $\frac{1}{2N}[\nabla_a M_b + \nabla_b M_a]$ in
the rest of the equations, so we obtain
\begin{equation}
R + \frac{1}{N^2} (\nabla \cdot M)^2 - \frac{1}{N^2} \nabla^a M_b
\nabla^b M^a = 16 \pi [(U^\bot)^2 \rho - p]
\end{equation}
\begin{equation}
\nabla_c\left[ \frac{1}{2N} (\nabla^c M_b + \nabla_b M^c)\right] -
\nabla_b(\nabla \cdot M) = 8 \pi \rho U_b U^\bot
\end{equation}
and
\begin{eqnarray}
R^a_b + \frac{1}{2N}(\nabla M)[\nabla^a M_b + \nabla_b M^a] &-&
\frac{\nabla^a \nabla_b N}{N} + \frac{1}{N} \mathcal{L}_M
[\nabla_b M^a +
\nabla^a M_b]\nonumber \\
   &=& \frac{1}{2} p \delta^a_b -
\rho [U^a U_b + \frac{1}{2}\; \delta^a_b(U^\bot)^2]
\end{eqnarray}
which are to be solved for $N, M^a$ and $\gamma_{ab}$.  The matter
evolution for WAT-III is governed by (11)-(12) and then on each time
slice we solve the ten equations (24)-(26) for the ten gravitational
variables.

\indent The gravitational equations for WAT-III and for WAT-II are
just about equal in complexity.  How is one to choose between
these two approximation schemes?  WAT-II, we have seen, is
motivated by the relationship between gravitational radiation and
the conformal geometry on maximal slices.  WAT-III, on the other
hand, is the natural generalization of the Bardeen-Wagoner
approximation treatment of axially symmetric systems [16].

\indent But the motivation for the various versions of WAT must be
only a secondary consideration.  These waveless theories are
proposed as possible tools for the approximate calculation of the
motion of multibody physical systems; they are \underline{not} to
be treated as possible candidates for the ``true theory of
gravity.''  [Like the Newtonian theory of gravity, all versions of
WAT involve action-at-a-distance and therefore violate any
possible notion of causality.]  So the various versions of WAT
 should be judged using more practical considerations primarily:
 ease of application, accuracy of approximation.  Such practical
 considerations cannot yet be properly judged.  More experience is
 needed in applying WAT-I, WAT-II (coupled or uncoupled), WAT-III,
 or any other version, to physical systems of interest.
 \pagebreak
{}

\begin{thebibliography}{16}
\bibitem[4]{} For definiteness, we consider our ``multibody
               system'' to be one which is filled with perfect fluid, satisfying
               Euler's equations of motion.  Our discussion here would change
               very little if we were to consider instead a system of discrete
               particles.
\bibitem[8]{} As in the Newtonian discussion, we choose perfect
              fluid sources here simply as an example.  WAT-I works with
              \underline{any} desired source field (as represented in $L_M$).
\bibitem[10]{} Existence and uniqueness proofs have so far been
               carried through only for systems with spherical or planar
               symmetry.
\end{thebibliography}
\end{document}